.

—

**A new approach to extracting coronary arteries and detecting stenosis in invasive coronary angiograms**

Chen Zhao [1], Haipeng Tang [2], Daniel McGonigle [2], Zhuo He [1], Chaoyang Zhang [2], Yu-Ping Wang [3], Hong-Wen Deng [3], Robert Bober [4]*, Weihua Zhou [1]*

.

1. Department of Applied Computing, Michigan Technological University, Houghton, MI, 49931, USA
2. School of Computing Sciences and Computer Engineering, University of Southern Mississippi, Hattiesburg, MS, 39406, USA
3. Tulane Center of Bioinformatics and Genomics, Tulane University School of Public Health and Tropical Medicine, New Orleans, LA, 70112, USA
4. Department of Cardiology, Ochsner Medical Center, New Orleans, LA, 70121, USA

[#]Authors for Correspondence:

Weihua Zhou, PhD

Department of Applied Computing, Michigan Technological University, Houghton, MI, 49931, USA

E-Mail: whzhou@mtu.edu

Or

Robert Bober, MD

Department of Cardiology, Ochsner Medical Center, New Orleans, LA, 70121, USA

Email: rbober@ochsner.org

**Abstract**

***Background***: In stable coronary artery disease (CAD), reduction in mortality and/or myocardial infarction with revascularization over medical therapy has not been reliably achieved. Coronary arteries are usually extracted to perform stenosis detection. As such, accurate segmentation of vascular structures and quantification of coronary arterial stenosis in invasive coronary angiograms (ICA) is necessary. However, performing accurate arterial segmentation and stenosis detection automatically remains a great challenge because of confounding factors, low contrast and moving frames in ICA.

***Purpose:*** We aim to develop an automatic algorithm by deep learning to extract coronary arteries from ICAs. Additionally, an automatic quantification and detection algorithm will be developed to analyze the stenosis in the coronary arterial tree.

***Methods:*** In this study, a multi-input and multi-scale (MIMS) U-Net with a two-stage recurrent training strategy was proposed for the automatic vessel segmentation. Incorporating features such as the Inception residual module with depth-wise separable convolutional layers, the proposed model generated a refined prediction map with the following two training stages: (i) Stage I coarsely segmented the major coronary arteries from pre-processed single-channel ICAs and generated the probability map of vessels; (ii) during the Stage II, a three-channel image consisting of the original preprocessed image, a generated probability map, and an edge-enhanced image generated from the preprocessed image was fed to the proposed MIMS U-Net to produce the final segmentation probability map. During the training stage, the probability maps were iteratively and recurrently updated by feeding into the neural network. After segmentation, an arterial stenosis detection algorithm was developed to extract vascular centerlines and calculate arterial diameters to evaluate stenotic level.

***Results and Conclusions:*** Experimental results demonstrated that the proposed method achieved an average Dice score of 0.8329, an average sensitivity of 0.8281, and an average specificity of 0.9979 in our dataset with 294 ICAs obtained from 73 patient. Moreover, our stenosis detection algorithm achieved a true positive rate of 0.6668 and a positive predictive value of 0.7043. Of note, our large image dataset covers the most commonly used view angles in clinical practice of ICA. Accordingly, our proposed approach has great promise to clinical use and could help physicians improve diagnosis and therapeutic decisions for CAD.

## 1. Introduction

Invasive treatment of stable coronary artery disease (CAD) either by percutaneous coronary intervention (PCI) or coronary artery bypass grafting (CABG) is common with over 800,000 procedures performed in the United States annually [1, 2]. The invasive coronary angiography (ICA) provides the gold standard for visualizing the anatomical structure of the coronary arteries and for further treatment [3]. ICA is a procedure where catheters are placed at the ostia of the major epicardial vessels and injected with contrast media. The contrast fills the lumen of the epicardial vessels in order to delineate coronary artery anatomy. Typically, this procedure is performed using X-ray fluoroscopy and images are recorded in 2D and in real time. The treatment decision is usually made based on visual assessment of CAD with percent stenosis being the most common metric used. However, identification of the coronary anatomy and their complete structures is challenging due to the following: 1) very low contrast, which may be caused by limited radiation dosage or light exposure; 2) moving objects; 3) ambiguities by a limited number of view angles and diluting contrast agent. Examples of ICA images and the corresponding vessels (used as the ground truth in this paper) are shown in Figure 1.

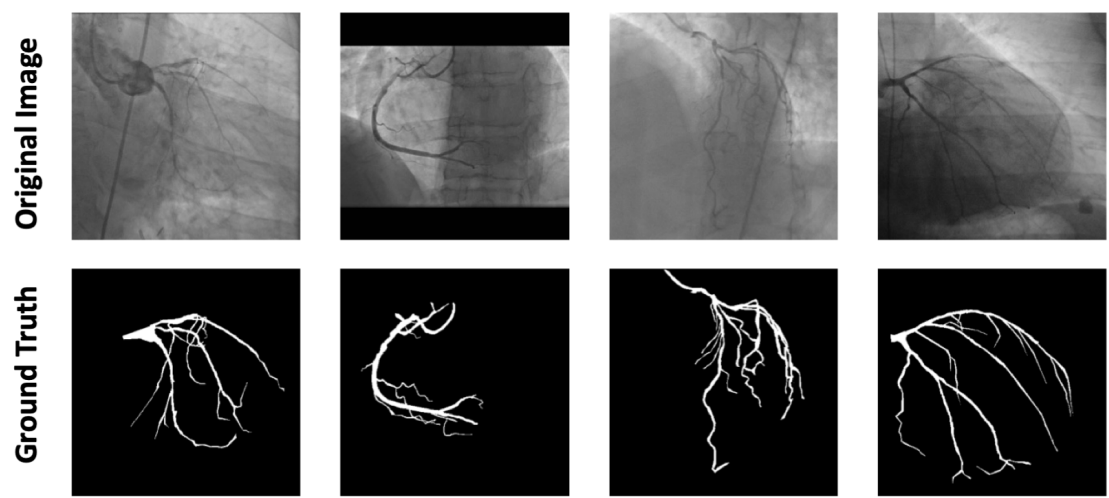

**Figure 1.** A close-up of invasive coronary angiograms and corresponding arteries. These arteries were manually extracted by experienced operators and used as the ground truth in this paper.

Automatic vessel segmentation of the ICA has drawn great attention of researchers and a number of segmentation methods have been proposed over the past decades. Existing vessel segmentation approaches based on traditional image processing can be divided into three categories: image filter-based methods, tracking-based methods, and model-based methods [4]. For the image filter-based segmentation approaches, wavelets [5], Gaussian filters [6] and Gabor filters [7] are used to extract low-level features and then a threshold-based classifier is adopted to generate final segmentation labels. For the tracking-based method, a set of initial pixels are selected, and the vessel tree structure grows according to specific rules, such as a brightness selection rule [8], expert tracing [9], etc. As for the model-based method, an active contours approach is widely used, including edge-based and region-based methods [10]. The methodology based on traditional image processing is not robust enough, especially when working on the low-contrast fluoroscopy angiograms.

Deep learning has demonstrated itself a powerful approach for a large number of computer vision tasks.

By using several convolutional layers and pooling layers, the hierarchical features of the images are extracted automatically, and then used for image recognition and segmentation tasks. Nasr-Esfahani et al. proposed a convolutional neural network (CNN) based method to predict the category of the center pixel in the patch and then to perform vessel extraction in X-ray angiograms [11]. They proposed another approach which used two CNNs sequentially to perform vessel segmentation from coarse to fine-grained [12]. Though existing approaches obtained impressive results, it is desired to further improve the model performance for vessel extraction to meet the clinical requirements.

In this paper, we firstly presented a new two-stage approach based on multi-input multi-scale U-Net (MIMS U-Net) for fully automatic vessel segmentation in ICAs: Stage I coarsely segmented the major coronary arteries from the pre-processed single-channel ICAs and generated the probability map of vessels; the vessel probability maps were then extracted and refined during the training iterations in Stage II. Experimental results show that the proposed MIMS U-Net achieved an average Dice score of 0.8329. After segmentation, an automatic stenotic quantification and detection algorithm was proposed. The vascular tree was divided into vessel segments and the most stenotic region within each arterial segment was used to evaluate the stenotic level. Considering the discrepancy of the stenosis levels between the detected stenotic points and the corresponding ground truth, the designed approach achieved a true positive rate of 0.6668 and a positive predictive value of 0.7043.

The innovations of the paper are as follows: we developed a MIMS U-Net with a two-stage recurrent training strategy for the vessel extraction. The MIMS U-Net contains three branches of the images with different resolutions to prevent feature loss. During the Stage II training procedure, the shape prior generated in Stage I is used to fine-tune the model weights and further enhance the precision of the segmentation results. In addition, the designed local Otsu binarization method converts the probability maps to binary images. Overall, our approach has great promise to advance to clinical use.

## 2. Materials and Methods

Figure 2 illustrates the workflow of our proposed approach for artery segmentation and stenosis detection. A deep learning-based method (MIMS U-net) was developed to extract coronary arteries in ICAs. Following are extracting centerlines of arterial segments and calculating the vessel diameters. Stenotic levels are thus assessed based on the diameters.

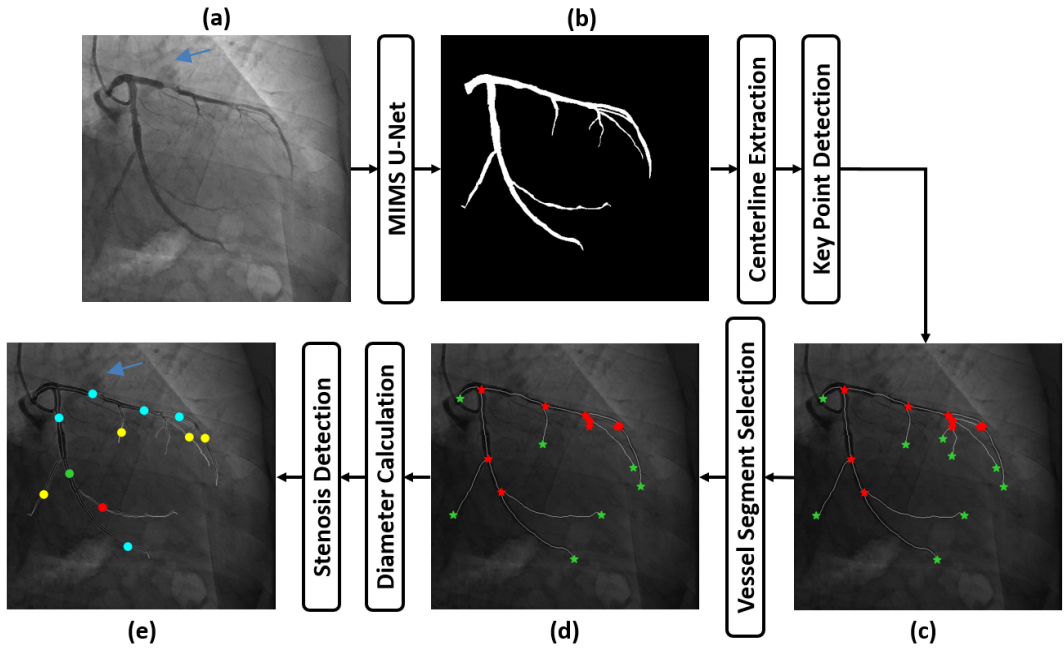

**Figure 2.** Workflow of our proposed approach for arterial segmentation and stenosis detection in invasive coronary angiograms. (a) Original image; (b) Segmentation result; (c) Arterial centerlines with detected key points. The red stars are joint points and the green stars are end points. (d) Centerlines of measured arterial segments. Capillaries whose diameters are smaller than 1.8mm are removed. (e) Detected stenosis. The points are detected stenotic points and the corresponding colors indicate different stenotic levels. Green, minimal stenosis; light green, mild stenosis; yellow, moderate stenosis; red, severe stenosis.

## 2.A. Invasive coronary angiogram acquisition and description

This retrospective study enrolled 73 patients who received ICA between September 2010 and November 2017. Invasive coronary angiography was performed using a Toshiba Infinix angiography system (Toshiba Medical Systems, Nasushiobara, Japan) and acquired at 15 frames/sec and tube voltage of 80kV. Each image has a size of 512×512 pixels. The pixel sizes range from 0.2442 mm to 0.3667 mm.

The dataset consists of 294 ICAs with 105 left coronary artery (LCA) images and 115 right coronary artery (RCA) images. For each patient, at least three standard projection views were captured. For each projection view, a frame which was used for anatomical structure analysis in clinical practice was selected for segmentation. Thus, the similarity between images is low. Table 1 illustrates the frequency of standard ICA views used for segmentation. The vessel contours were manually drawn by well-trained operators, confirmed by an experienced interventional cardiologist and then provided to this study as the ground truth.

**Table 1.** Frequency of standard ICA views. LCA, left coronary artery; RCA, right coronary artery. LAO, left anterior oblique; RAO, right anterior oblique; AP, antero-posterior; CRA, cranial; CAU, caudal.

|     | LAO+CRA | LAO+CAU | RAO+CAU | RAO+CRA | AP | Total |
| --- | --- | --- | --- | --- | --- | --- |
| LCA | 61 | 8 | 48 | 61 | 1 | 179 |
| RCA | 46 | 24 | 21 | 24 | 0 | 115 |

## 2.B. Image preprocessing

The top-hat and N4 bias correlation are sequentially applied to enhance the contrast of fluoroscopy images. In order to increase the number of training images, data augmentation is also used by randomly flipping the images around vertical and horizontal axes and randomly rotating from -30° to 30°. Due to the GPU memory limitation, we cropped patches from the preprocessed image with a fixed size of 384 $\times$ 384 for model training. To prevent information loss, the patches are cropped using a window with the size of 384$\times$384 and sliding through both the horizontal and vertical axes at a step of 32.

## 2.C. Multi-input multi-scale U-Net

In this section, we propose a MIMS U-Net model for coronary vessel segmentation. First, we refine the original U-Net and significantly enhance the multi-scale capability. Secondly, we use a two-stage training strategy to fine-tune the network so that the segmentation results are updated from coarse-grained to fine-grained. Finally, a post-processing step, including patched Otsu [13] threshold and maximum connected region search, is employed to remove the noise and to generate the final segmentation results. The Otsu algorithm can calculate the optimum threshold to convert a grayscale image to a binary image.

The encoder of the original U-Net proposed by Ronneberger et al. [14] uses convolution layers to automatically extract features and max-pooling layers to down-sample feature maps. In the encoder part of U-Net, the number of feature maps increases with the decrement of the resolution. And the earlier layers of CNN learn low-level image features, such as shape, edge, etc., while the higher layers learn high-level features, which are more important to the specific application. The original U-Net crops the centers of feature maps in the encoder and then concatenates them with the feature maps in the up-sampling layer in the decoder. The cropping operation causes the model to lose some features, so researchers in [15] remove the cropping operation and directly feed the feature maps extracted from the encoder into the decoder by using the skip-connection. However, the existing U-Net models have two limitations: (*i*) to fit the GPU memory, increase the receptive field, extract hierarchal features, and keep the invariance of the CNN, pooling layers are always employed in the encoder for deepening the network, which will not preserve all the spatial information [16]. (*ii*) if the features are lost in shallow convolutional layers, then the deepening layers with low resolutions cannot recognize vessels with a large variety of diameters.

To address the problem (*i*), we modified the U-Net architecture as follows: we keep the extracted feature maps before each down-sampling layer and add multi-resolution branches to the neural network, as depicted in Figure 3.

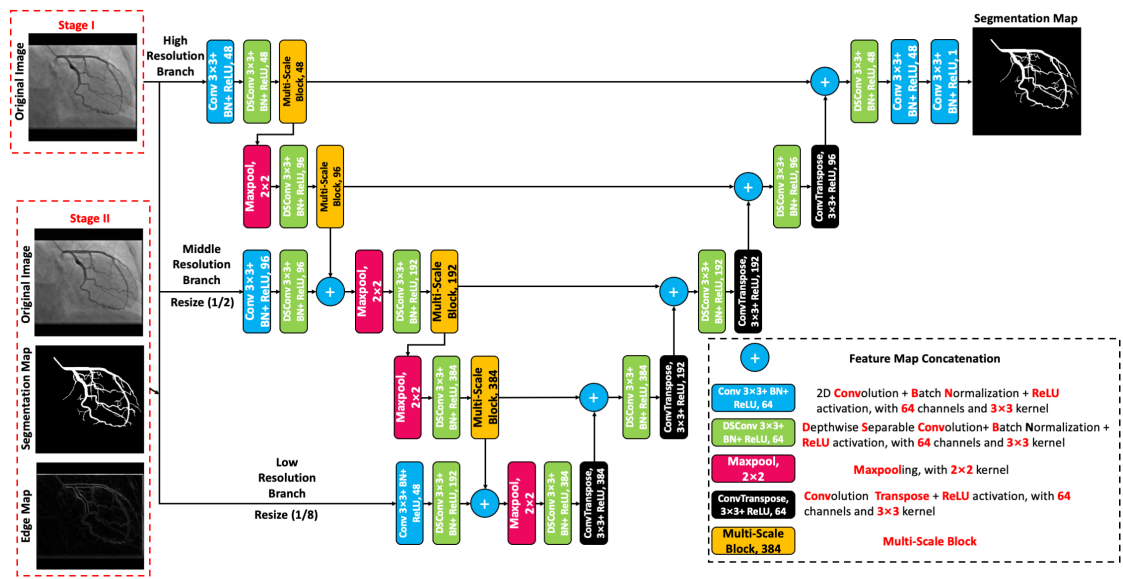

**Figure 3.** Flow diagram of our proposed multi-input multi-scale U-Net. Blocks with different colors represent different types of layers. The numbers in the blocks are the quantity of feature maps. In addition, the numbers in the parentheses are the zooming factors of input images.

To overcome the problem in (*ii*), we set three input branches, including a low-resolution branch, a middle-resolution branch, and a high-resolution branch, to help the U-Net learn both high-level features and raw information in fluoroscopic images.

**Low-resolution branch**: The input size of the low-resolution branch is only 1/8 of the original input; however, it is applied in the deepest structure of the network and contains the most layers which could efficiently extract the most semantic information, which is highly related to fine classification. The low-resolution branch is firstly fed into a conventional convolutional layer and a depth-wise separable convolution block to extract high-level features. The model then concatenates the multi-scale features from the top of the network to collaboratively encode the feature maps. In the vessel segmentation task, the low-resolution branch learns the high-level features for pixel classification.

**High- and middle-resolution branch:** The input size for the high- and middle-resolution branches are 1 and 1/2 of the original input. With small numbers of convolutional layers and feature maps, i.e. 48 and 96, the total number of parameters in such branches is constrained. However, only with a small number of convolutional layers (two layers), the two branches could learn shallow features, such as edges and shapes, and detect the major (larger) arteries in fluoroscopy angiograms. To prevent losing the features from the application of max-pooling layers [17], we add a middle-resolution branch after the high-resolution branch.

To enlarge the receptive field and address the problem (*ii*), we adopt a type of multi-scale convolutional block into the encoding path of proposed U-Net. From Figure 1, it is clear that the thickness of vessels varies in different sections of images. To improve our model performance, it is necessary to enhance the multi-scale capability of the neural networks. As illustrated in Figure 4, we employ four multi-scale blocks in different parts of the proposed U-Net. In this paper, we employ convolution blocks with different kernel size, such as $3 \times 3$, $5 \times 5$ and $7 \times 7$ to enhance multi-scale capacity. Inspired by the design of Inception module [18], the receptive field of a $n \times 1$ convolution followed by an $1 \times n$ convolution is equivalent to a $n \times n$ convolution layer. The number of parameters is reduced from $n^2$ to $2n-1$. Therefore, we factorize each convolutional layer into asymmetric convolutions to reduce the number of parameters. The architecture of the multi-scale block is shown in Figure 4.

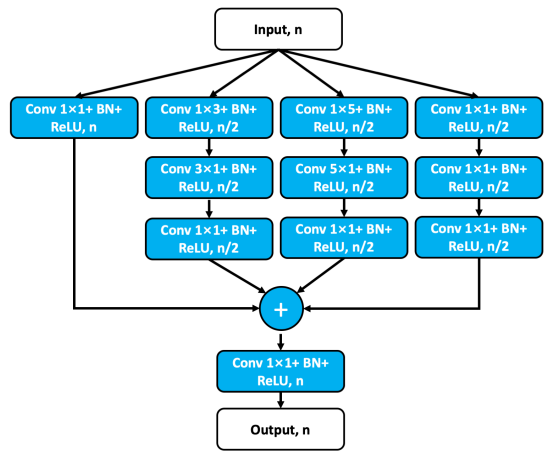

**Figure 4.** Architecture of the multi-scale block.

## 2.D. Multi-stage training

In order to segment the vessel tree from the image background, the pre-processed grayscale images are fed into the proposed U-Net. Due to the GPU memory limit, the original images are cropped with a patch size of $384 \times 384$ and the U-Net generates the probability map with the same resolution. The output of Stage I could capture the vessel regions with a relatively high precision, but it is difficult to detect the relatively small vessels. Due to the variation of grayscale transformation in major arteries, Stage I often misses such regions, which significantly decreases the model performance.

In order to resolve these problems, we design a two-stage training strategy to refine the vessel boundary and capture the regions in major arteries in the vessel tree. The workflow of the proposed approach is shown in Figure 5.

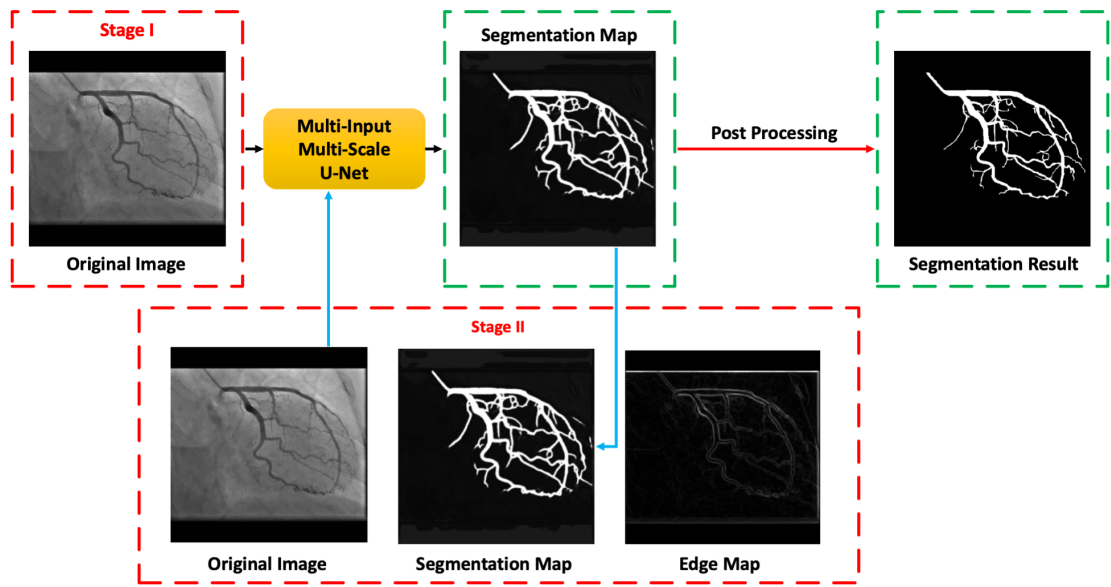

**Figure 5.** Overview of our proposed method for coronary artery segmentation.

As shown in Figure 5, the model input in Stage II incorporates three channels, including a pre-processed image with high contrast, generated segmentation map and denoised edge-enhanced map:

(1) Pre-processed image with high contrast. By applying an N4 bias correlation [19] and top-hat transform, the contrast of the original image is enhanced.

(2) Generated segmentation map. The probability map contains the most information for the final segmentation and provides guidance for sequential training. To accelerate the model training, we feed the previously generated segmentation map as a channel of input into the proposed U-Net in Stage II. The probability maps are updated during the training process, and thus we dynamically change the model input in each training iteration.

(3) Denoised edge-enhanced map. Edge enhancement will facilitate the model to capture the vessels. The edge of the vessel can be defined as the prior knowledge for model training. We use a canny edge detection algorithm to generate the edge maps of the enhanced images. To remove the noise in edge maps, a Gaussian filter is adopted to generate the final edge map.

Then, we stack the three types of images into a three-channel image as the model input in Stage II. The difference of the inputs between Stage I and Stage II is whether it contains the generated edge map and the generated probability map or not.

Admittedly, we could feed the edge map into Stage I for model training. However, experimental results demonstrate that detecting vessel boundary at the first training epochs is difficult and often causes the model not to converge. As for the probability map, its accuracy in Stage I could not be guaranteed because the parameters in U-Net are randomly initialized and there exists a great amount of noise. While in Stage II, the probability map is relatively precise and is beneficial for model training. Thus, Stage I contains a single-channel input while the Stage II contains a three-channel input. By iteratively training the model, the vessel map is extracted from coarse-grained to fine-grained. Technically, we stacked three one-channel gray scale images as the input of Stage I training. In addition, the edge map, segmentation map and original image were stacked as a three-channel input in Stage II. Hence, the model inputs in both Stage I and II are three channel images so that the weights generated at the end of Stage I can be used as the initial weights for Stage II training. To predict a new set of data, firstly we need to feed the images into the model to generate the corresponding edge map and probability map by using the weights trained at the end of Stage I; then we stack them as three-channel images to generate final segmentation results with the parameters in Stage II.

### 2.E. Post-processing

As described in Section 2.C and 2.D, the U-Net segmentation network generates a corresponding probability map, while not a binary segmentation mask. A post-processing step is applied to obtain the binary segmentation result. A patched threshold method based on Otsu is used. The size of the original image is $512 \times 512$. Our patched threshold method uses a window size of $384 \times 384$ to slide through both the horizontal and vertical axes at a step of 32. Each scanned patch is processed by the Otsu algorithm. As for the pixel located in the overlapping region among the sliding windows, it was assigned as a foreground pixel (vessel) if it was classified to 1 in any Otsu binarization process. Finally, we stitch all binary patches together to generate segmentation results with the original image resolution.

Since the coronary vessels have treelike anatomy and are fully-connected to each other, we select the maximum connected region from the predicted binary image to remove disconnected "noise branches" and generate final segmentation results.

### 2.F. MIMS U-Net optimization strategy

In image segmentation, a pixel-wise loss function is usually used to penalize the difference between the ground truth and the predicted probability map [20]. The pixel-wise loss function is commonly defined by a cross-entropy as follows,

$$L_{pixel-wise} = \sum_i -y_i \log(\hat{y}_i) - (1-y_i) \log(1-\hat{y}_i), \tag{1}$$

where $y_i$ and $\hat{y}_i$ represent the ground truth and the predicted probability map of pixel $i$, respectively. The ground truth is a binary image, which means that '1' represents the vessel and '0' indicates the background. In addition, we add L2 regularizer to our loss function to regularize the model. Finally, our total loss function for the segmentation is defined as a linear combination of these two loss functions mentioned above,

$$L_{loss} = L_{pixel-wise} + L_2 \tag{2}$$

To prevent over-fitting, we add a Dropout layer before the last convolutional layer and set the dropout probability of 0.8.

### 2.G. Stenosis detection

To perform the stenosis detection, the fundamental task is to extract vascular centerlines and calculate variation of the diameters along the vessel segment. The preserved pixels along the centerlines are located at the middle of the vessel segments and the distance between the boundaries of the vessel contour is defined as the corresponding pixel-wise diameter.

To extract arterial centerlines, we utilize erosion to iteratively shrink the vascular tree until the its topology remains unchanged [21]. Then, a Euclidean distance transform method is used to calculate the vessel diameters [22]; thus, each pixel in the vascular centerline is assigned with a diameter. After obtaining the arterial centerlines and the diameters, a stenosis detection algorithm is developed. First, we use an edge linking algorithm [23] to detect connected edge points. According to the detected joint points and end points, we separate the entire vascular tree into arterial segments. For each arterial segment, the maximal diameter is defined as the reference diameter, which is denoted as $d_{max}$. The minimal diameter of the entire arterial segment is defined as the minimal diameter, which is denoted as $d_{min}$. The stenotic level of the vessel segment is defined in Eq. 3.

$$s = \left(1 - \left(\frac{d_{min}}{d_{max}}\right)\right) \times 100\% \tag{3}$$

The stenotic levels are reported for only the major vessel segments with the maximal diameter ≥1.8 mm. In other words, the vessel segments with the maximal diameter <1.8 mm were ignored. Note that the pixel number which is are equivalent to 1.8 mm varies with the pixel size in a specific image.

The stenotic levels are categorized into minimal, mild, moderate and severe with the stenosis percent ranges of (10%, 24%), (25%, 49%), (50%, 69%) and (70%, 100%), respectively [24].

## 2.H. Experimental Settings

Our approach was implemented in Python by TensorFlow 1.13. We trained the deep learning model through Adam optimizer [25] with a base learning rate of 0.0001 and a decay rate of 0.05. We deployed our model on a workstation with a Titan V100 GPU with 16GB GPU memory. In our MIMS model, the batch size could only be set as 1 due to the GPU limit. Each training iteration took about 4 minutes and the model converged within 300 and 200 epochs in Stage I and Stage II, respectively.

We used 5-fold cross-validation to evaluate the model performance. For each fold, 235 angiograms (80%) were used for training, and the rest (20%) were for the testing.

## 3. Experiments and Results

### 3.A. Evaluation Metrices

To quantitatively reflect the performance of our segmentation model, we use the Dice score, sensitivity (SN) and specificity (SP) to evaluate the experimental results based on pixel information. The true positive (TP) pixel indicates that the pixel belongs to a vessel and our model predicts the pixel as a vessel pixel; the false positive (FP) pixel denotes that it is a background pixel but predicted as a vessel pixel. The true negative (TN) and false negative (TP) can be defined as a similar manner. Therefore, the SN and SP are defined as:

$$SN = \frac{TP}{TP+FN} \tag{4}$$

$$SP = \frac{TN}{TN+FP} \tag{5}$$

Moreover, the stenotic levels by our approach are validated against those from the ground truth. If a stenotic point is detected in the predicted arterial tree and is matched with the ground truth, then the point is denoted as a true positive point (TPP). If a stenotic point is detected in the predicted arterial tree and there is no matched point in the ground truth, then it is denoted as a false positive point (FPP). On the contrary, if the stenotic point from the ground truth is not found in the model prediction, then it is defined as a false negative point (FNP). Consequently, the following true positive rate (TPR) and positive predictive value (PPV) are calculated to evaluate stenosis detection by our approach, as shown in Eq. 6 and Eq. 7.

$$TPR = \frac{TPP}{TPP + FNP} \tag{6}$$

$$PPV = \frac{TPP}{TPP + FPP} \tag{7}$$

Furthermore, root mean squared error (RMSE) is calculated to quantitively reflect the discrepancy of stenotic levels between our approach and the ground truth. The RMSE is defined in Eq. 8.

$$RMSE = \sqrt[2]{\frac{1}{N} \sum_{1}^{n} (b_e - b_g)^2} \tag{8}$$

where $b_e$ is the estimated stenotic level by our approach, $b_g$ is the ground truth and $N$ is the number of the TPPs.

### 3.B. Experimental results of artery segmentation

To evaluate the effectiveness of our proposed model, we set two baseline models by separately removing the multi-scale block and multi-input block and then generate a Multi-Input U-Net (MI U-Net) and a Multi-Scale U-Net (MS U-Net). In the MI U-Net, the multi-scale blocks in MIMS U-Net are replaced by depth-wise separable convolutions with a kernel size of 3×3. For the MS U-Net, the middle and low-resolution branches are removed. Both the MI U-Net and MS U-Net were trained with the loss function defined in Eq.2. Examples of resulting vessels are delineated in green, as illustrated in Figure 6.

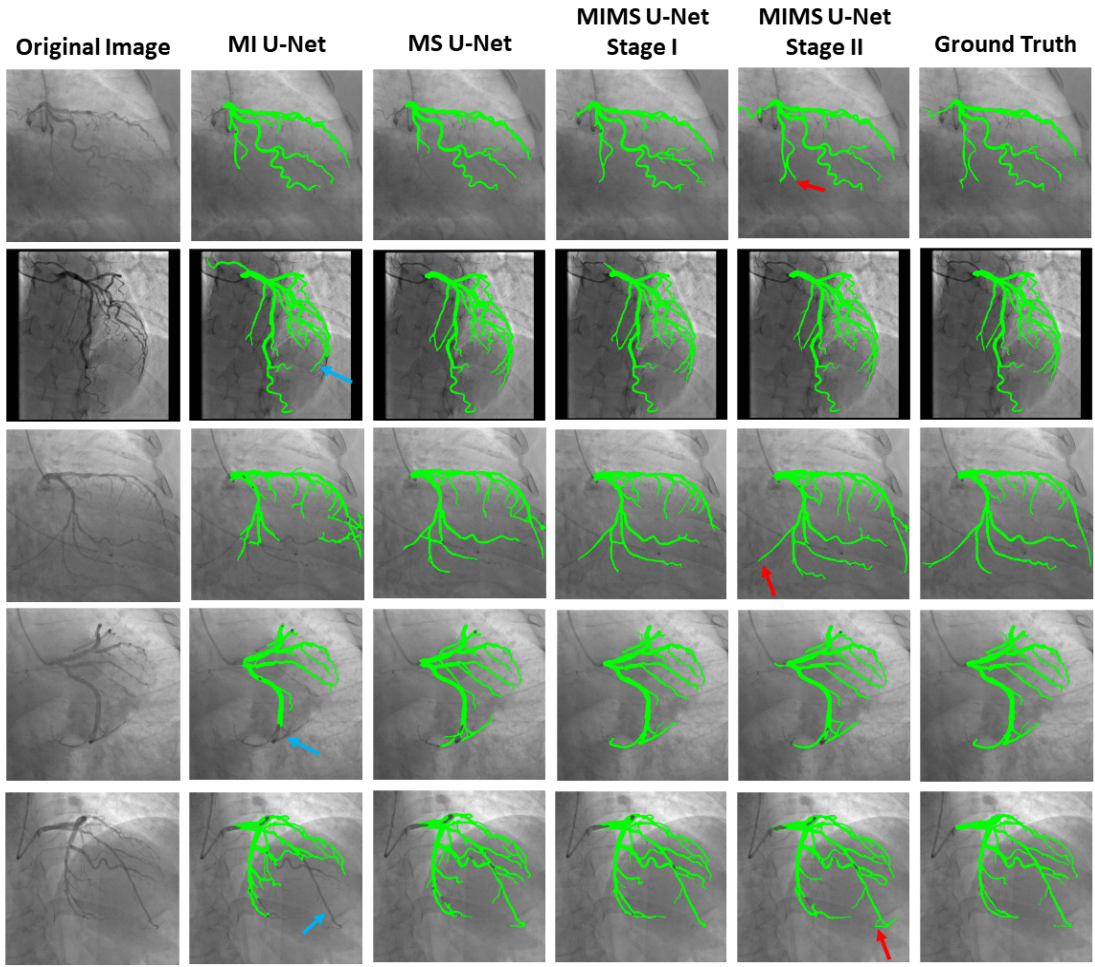

**Figure 6**. Illustrations of segmentation results obtained by different methods.

The comparison of the model performance is given in Table 2.

**Table 2** Comparison of the proposed method with baseline models. "no" represents that the proposed post-processing was not used, and the binarization was addressed by Otsu algorithm; "yes" represents that the proposed post-processing was used.

| Methods | Post-processing | SN | SP | Dice |
|---|---|---|---|---|
| MI U-Net | no | 0.7709±0.1188 | 0.9897±0.0055 | 0.7392±0.1288 |
| MI U-Net | yes | 0.7725±0.0492 | 0.9891±0.0023 | 0.7597±0.0327 |
| MS U-Net | no | 0.7780±0.0868 | 0.9894±0.0062 | 0.7657±0.1134 |
| MS U-Net | yes | 0.7892±0.0101 | 0.9918±0.0015 | 0.7946±0.0243 |
| MIMS U-Net (without Stage II) | no | 0.8371±0.0831 | 0.9911±0.0053 | 0.8062±0.0699 |
| MIMS U-Net (without Stage II) | yes | 0.8102±0.0301 | 0.9935±0.0002 | 0.8072±0.0034 |
| MIMS U-Net (with Stage II) | no | 0.8217±0.1029 | 0.9913±0.0726 | 0.8299±0.0914 |
| MIMS U-Net (with Stage II) | yes | 0.8281±0.0301 | **0.9979±0.0001** | **0.8329±0.0021** |

From Table 2, we can observe that with the proposed post-processing step and the Stage II training procedure, the MIMS U-Net achieved a higher SP and Dice score.

To illustrate the effectiveness of the modified local thresholding algorithm for post-processing, several image binarization methods were applied to the MIMS U-Net generated probability maps. We implemented the thresholding methods [13, 26-29] to binarize the probability maps and to evaluate the model performance. The comparison of the thresholding algorithms is shown in Table 3.

**Table 3.** Comparison of the different binarization techniques and settings to convert the probability maps from our proposed method into the binary segmentation masks. Full binarization indicates that the binarization used the full-size probability maps; and patched binarization represents the binarization used the patched probability maps. Numbers in the parentheses shows the corresponding patch size.

| Methods | Type | SN | SP | Dice |
|---------|------|----|----|------|
| Li et. al. [26] | Full | 0.8210±0.1254 | 0.9913±0.0044 | 0.8291±0.0884 |
| Yen et. al. [27] | Full | 0.9175±0.0894 | 0.8612±0.3893 | 0.7510±0.0901 |
| Zack et. al. [28] | Full | 0.9101±0.0710 | 0.9533±0.0615 | 0.7861±0.1242 |
| Sauvola et. al. [29] | Patched (15) | 0.7892±0.0010 | 0.9918±0.0015 | 0.7946±0.0243 |
| Otsu [13] | Full | 0.8217±0.1029 | 0.9913±0.0726 | 0.8299±0.0914 |
| Patched Otsu | Patched (128) | 0.8237±0.0024 | 0.9978±0.0001 | 0.8316±0.0109 |
| Patched Otsu | Patched (384) | 0.8281±0.0301 | **0.9979±0.0001** | **0.8329±0.0021** |

According to Table 3, our developed patched Otsu outperformed other image binarization algorithms in SP and Dice.

### 3.C. Experimental results of stenosis detection

Figure 7 shows example results of arterial segmentation and stenosis detection.

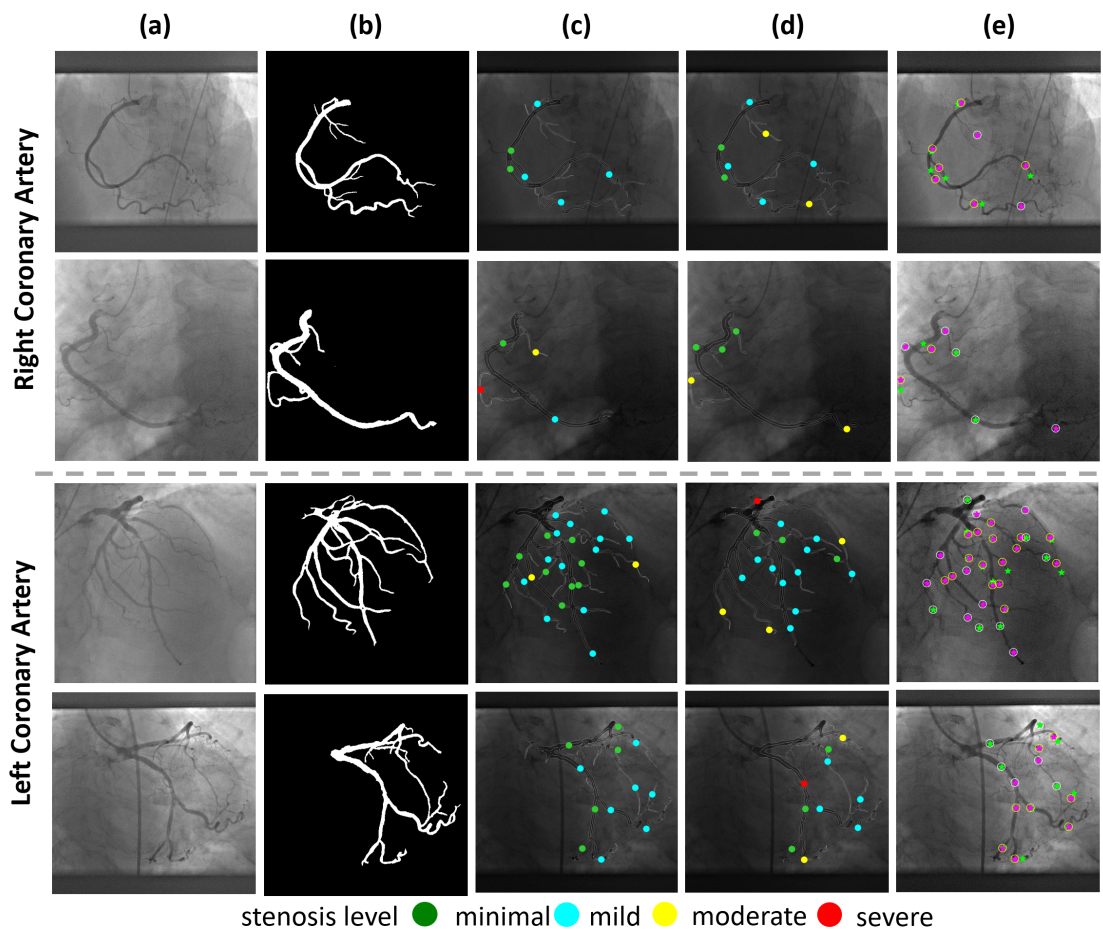

**Figure 7.** Examples of arterial segmentation and stenosis detection. Top: results using ICAs for right coronary artery; bottom: results using ICAs for left coronary artery. (a) original ICAs; (b) Segmentation results using our MIMS U-Net; (c) Stenosis detection on predicted arteries; (d) Stenosis detection on the arterial ground truth; (e) Correspondence of the stenotic levels between our approach and the ground truth. In (c) and (d), different colors represent different levels of stenotic levels. The pink and light green stars represent the stenotic points detected from the ground truth and the model prediction. The pink stars with yellow circle indicate true positive points; the pink stars with white circle are false negative points; and the light green stars with white circle are false positive points.

We evaluated the TPR, PPV and RMSE defined in section 3.A to reflect the performance of our entire approach. As a comparison, Table 4 shows the performance for MI U-Net, MS U-Net and MIMS U-Net.

**Table 4.** Comparison of the stenosis detection performance using arterial contours generated by MI U-Net, MS U-Net and MIMS U-Net. TPR, true positive rate; PPV, positive predictive value.

| Segmentation Model | Stenotic Level | TPR | PPV | RMSE | # TPP | # FNP | # FPP |
|---|---|---|---|---|---|---|---|
| MI U-Net | all | 0.5652 | 0.6481 | 0.1886 | 1811 | 1393 | 983 |
| | minimal | 0.5597 | 0.5744 | 0.1823 | 328 | 258 | 243 |
| | mild | 0.6058 | 0.6111 | 0.1335 | 844 | 549 | 537 |
| | moderate | 0.5452 | 0.7559 | 0.2251 | 536 | 447 | 173 |
| | severe | 0.4256 | 0.7744 | 0.3313 | 103 | 139 | 30 |
| MS U-Net | all | 0.6476 | 0.6318 | 0.1754 | 2518 | 1370 | 1467 |
| | minimal | 0.6666 | 0.5980 | 0.1573 | 500 | 250 | 336 |
| | mild | 0.6900 | 0.6174 | 0.1285 | 1133 | 509 | 702 |
| | moderate | 0.6096 | 0.6488 | 0.2099 | 717 | 459 | 388 |
| | severe | 0.5250 | 0.8038 | 0.2965 | 168 | 152 | 41 |
| MIMS U-Net | all | **0.6668** | **0.7043** | **0.1742** | 2142 | 1070 | 899 |
| | minimal | 0.6535 | 0.6278 | 0.1767 | 383 | 203 | 227 |
| | mild | 0.6961 | 0.7155 | 0.1279 | 976 | 426 | 388 |
| | moderate | 0.6554 | 0.7265 | 0.2063 | 643 | 338 | 242 |
| | severe | 0.5761 | 0.7692 | 0.2637 | 140 | 103 | 42 |

According to Table 4, the stenosis detection algorithm identified 2735 TPPs, 653 FNPs and 696 FPPs using the arterial contours generated by the proposed MIMS U-Net. The algorithm achieved a TPR of 0.6668, a PPV of 0.7043 and an RMSE of 0.1742 on all types of stenosis, which outperformed the results obtained by using the contours from MI U-Net and MS U-Net.

## 4. Discussion

### 4.A. Performance analysis for coronary artery segmentation

In Figure 8, we juxtapose the original vessel grayscale images, the segmentation results obtained by the original U-Net, by MIMS U-Net with Stage I training and Stage II training, and the corresponding ground truth images with overlay. Of note, the sizes of coronary vessels are different as shown in Figure 7; this is a key reason that the MI U-Net without the multi-scale technique fails to effectively distinguish the major arteries and capillaries from the background, as annotated with the blue arrows in Figure 6. Our multi-input and multi-scale techniques and two-stage training strategy in the MIMS U-Net significantly mitigate this problem, as annotated with red arrows in Figure 8.

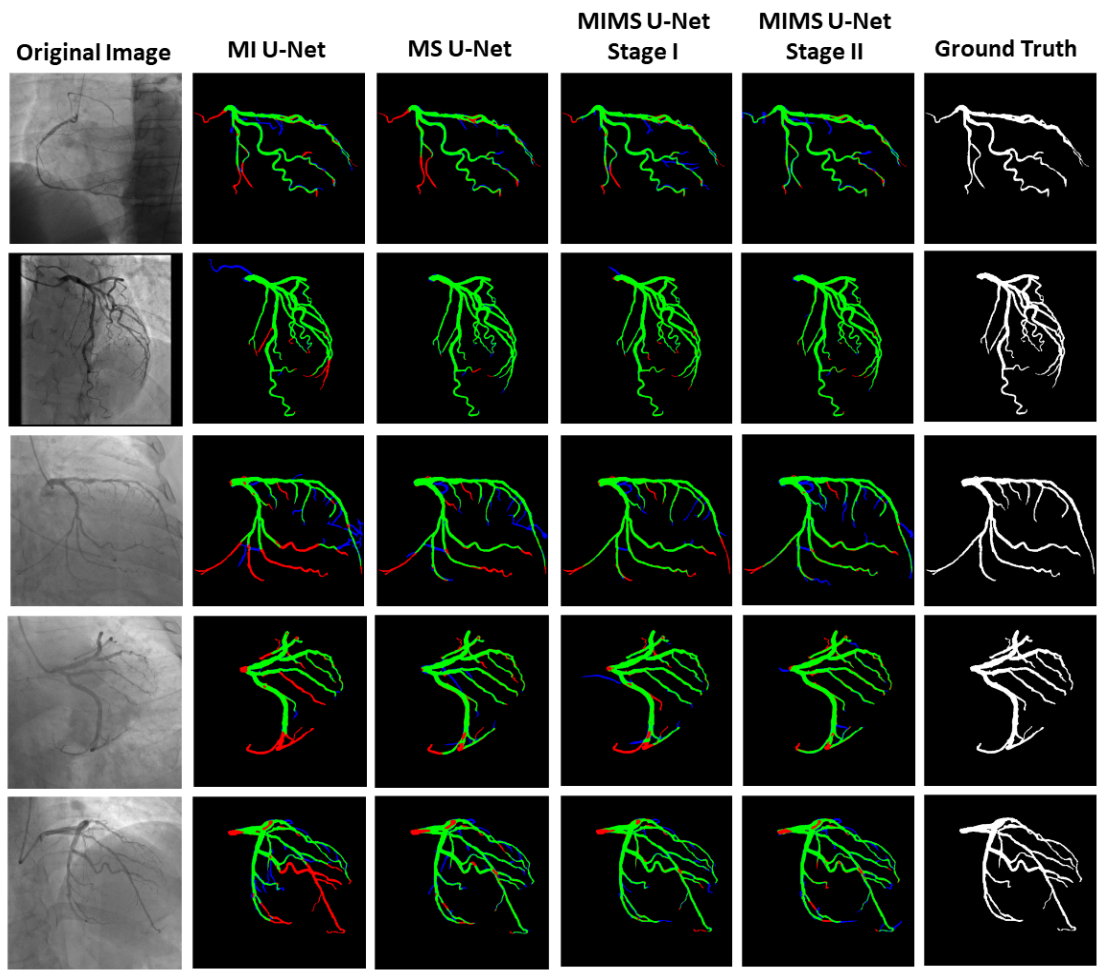

**Figure 8.** Difference maps of segmentation results against the ground truth. The green, red and blue regions represent true positive (TP), false negative (FN) and false positive (FP) pixels, respectively.

All the three models we develop (MI U-Net, MS U-Net, and MIMS U-Net) are evaluated using 5-fold cross-validation. From Table 2, MIMS U-Net outperformed the MI U-Net and MS U-Net in SN, SP and Dice score. Our proposed model with post-processing and Stage II training outperformed the other models in the SP and Dice score. Furthermore, by using the Stage II fine-tuning procedure and the designed post-processing algorithm, the Dice score of MIMS U-Net increased from 0.8062 to 0.8329.

The pooling layers in the CNN are beneficial to capture the hierarchal features and keep the invariance of CNN. However, they do not preserve all the spatial information during the feature map reduction process. For example, max-pooling chooses the max value of the grid, but the choice is not the true maximum in many situations [30]. Moreover, looking from an image downscaling perspective, max-pooling leads to implausible looking results [17]. We tried to mitigate the feature loss problem during the pooling process by adding the multiple inputs into the networks, and designed three branches for the MIMS U-Net presented in Section 2.C. In addition to feeding the down-sampled feature maps into the CNN, we input the feature maps of resized images into the middle-resolution branch and low-resolution branch of the designed MIMS U-Net. Experimental results in Section 3.B demonstrated that the MIMS U-Net achieved higher SN, SP and Dice score in our coronary artery segmentation task, compared to MS

U-Net which doesn't have multiple inputs.

Similar studies using deep learning for coronary vessel segmentation have been performed. However, to the best of our knowledge, only two studies are found and added for the comparison. Because each study has its own image dataset, and we don't have their source codes, the performance comparison in Table 5 can only provide a limited justification. Nevertheless, since our image dataset has a big sample size and covers the most commonly used view perspectives in clinical practice, the performance of our proposed method indicates great potential for clinical use.

**Table 5.** Comparison of the existing methods with our proposed method.

| Methods | Data | SN | SP | Dice |
|---------|------|-----|-----|------|
| CNN [11] | 44 X-ray angiography images | 0.7935 | 0.9890 | 0.7562 |
| CNN [12] | 44 X-ray angiography images | 0.8676 | 0.9859 | 0.8151 |
| MIMS U-Net | 294 ICAs | 0.8281±0.0301 | **0.9979±0.0001** | **0.8329±0.0021** |

In [11], the original vessel images were pre-processed by a top-hat transformation and the CNN network contained only 2 convolutional layers, which severely limits the model performance. Besides, the authors cropped the images into several patches with a resolution of 35×35. With a small patch size, it is difficult for convolution kernels to capture the global information and to generate a precise segmentation result. In [12], the authors implemented two CNNs. Both inputs were the patches and the outputs of them were just the central pixel of each patch. For each image, the model in [12] has to infer hundreds of patches in one angiogram to generate the final segmentation result. On the contrary, our model has a large receptive field and can generate the final segmentation result within two inference procedures.

Moreover, based on the difference maps, as illustrated in Figure 8, it is clear that the number of the FN pixels, shown in red, is less than the number of the FP pixels, shown in blue in our proposed MIMS U-Net model. The FN pixels often occur in capillary regions which are less important than major arteries in clinical practice. FN indicates the vessel pixels that are missed by our segmentation mask. If FN pixels occur frequently due to under-segmentation, doctors may be misled to characterize a patient as worse than conditions warrant. By contrast, FP pixels indicate false warning or over-segmentation by our model. Compared with the results generated by baseline models, the number of FN and FP pixels are smaller, which indicates that the proposed model has great advantages in clinical practice.

### 4.B. Performance analysis for stenosis detection

According to the results in Table 4, our stenosis detection approach with the MIMS U-Net segmentation achieved a TPR of 0.6668, a PPV of 0.7043 and an RMSE of 0.1742 on all types of stenosis. It outperformed other two models (MI U-Net and MS U-Net), suggesting the advantage of MIMS U-Net. Furthermore, the numbers of detected FPPs and FNPs by MIMS U-Net are significantly smaller than those by either MI U-Net or MS U-Net.

### 4.C. Application and clinical overview

Our approach achieved a Dice score at 0.8329 in arterial segmentation, and a TPR of 0.6668, a PPV of 0.7043 in stenosis detection. During the model training period, a powerful GPU is required. However, only a CPU is needed for prediction during test. For any new set of data, the images are fed into the trained model to generate the corresponding edge map and probability map. Afterwards, they are stacked as a three-channel image in the model to predict the final output. With an Intel Core I5 CPU, our model could extract vessel contours for a new image in 2.7 seconds after the model weights are loaded into the system memory, which is clinically acceptable. In addition, both the artery segmentation and stenosis detection are completely automatic. Accordingly, our approach has great promise for clinical use.

More importantly, our proposed approach brings new opportunities to further improve the diagnosis and treatment CAD. Myocardial revascularization (MR) is a standard interventional treatment for patients with stable CAD [31]. Most MR decision making is based on the percentage of stenosis. However, despite the frequency and commonality of MR procedures, reduction of the rates of mortality and/or myocardial infarction over medical therapy has not been realized, as summarized in the American College of Cardiology guidelines [32, 33]. To better visualize the anatomical structure of the coronary arteries, revascularization decisions are made at the time of performing ICA. The first requirement is that angiographic data be converted into 3D datasets. In addition, physiologic data, such as positron emission tomography (PET) stress testing, is performed as a gateway prior to anatomic testing by means of invasive coronary angiography. And at the time of clinical decisions for revascularization, anatomic testing (ICA) and physiologic testing (PET) are clinically segregated. Therefore, fusion of anatomic data from ICA with physiologic data at the level of point of care could improve outcomes. Thus, automatic extraction of coronary anatomy in this study is the initial step that is required for fusing angiographic data with functional data [34, 35]. The fused images, in addition to the original 2D dataset would be used for clinical decisions.

Nevertheless, there are several limitations in our approach. To train our model, we need to perform a two-stage training procedure which is time-consuming. Due to the use of multi-scale block in the designed U-Net, the model requires large GPU memory to train the model. In addition, the prediction time is longer compared to end-to-end one-step approaches.

In the future, we will further improve the performance of our segmentation model to generate more precise results. In addition, we will focus on the semantic segmentation of coronary vessels to extract different vessel segments: left anterior descending (LAD) artery, left circumflex artery (LCX), and etc. Then we will integrate coronary arteries from our approach with functional data to support clinical decision making for improved outcomes of myocardial revascularization.

## 5. Conclusion

In this paper, we developed and validated an automatic approach for arterial segmentation and stenosis detection in ICAs. A new two-stage MIMS U-Net model was proposed for arterial segmentation, which enlarges the receptive field with a multi-scale block and improves the performance by the capture of global features with the multi-channel input. The quantitative evaluation demonstrates the state-of-the-art performance of our approach. It has great promise to advance to clinical use.

**Acknowledgments**

This research was supported by a grant from the American Heart Association under award number 17AIREA33700016, a grant from Ochsner Hospital Foundation (PI: Weihua Zhou), and a new faculty startup grant from Michigan Technological University Institute of Computing and Cybersystems (PI: Weihua Zhou). This research was also supported in part by the National Institutes of Health under award numbers U19AG055373, R01GM109068, R01MH104680, R01MH107354 and by the National Science Foundation NSF under award number 1539067.

# References

1. Cram, P., et al., *Indications for percutaneous coronary interventions performed in US hospitals: a report from the NCDR®.* American heart journal, 2012. **163**(2): p. 214-221. e1, 0002-8703.

2. Melly, L., et al., *Fifty years of coronary artery bypass grafting.* Journal of thoracic disease, 2018. **10**(3): p. 1960.

3. Li, Z., et al., *A robust coronary artery identification and centerline extraction method in angiographies.* Biomedical Signal Processing and Control, 2015. **16**: p. 1-8, 1746-8094.

4. Tsai, Y.-C., H.-J. Lee, and M.Y.-C. Chen, *Automatic segmentation of vessels from angiogram sequences using adaptive feature transformation.* Computers in biology and medicine, 2015. **62**: p. 239-253, 0010-4825.

5. Bankhead, P., et al., *Fast retinal vessel detection and measurement using wavelets and edge location refinement.* PloS one, 2012. **7**(3): p. e32435, 1932-6203.

6. Li, Y., et al. *A novel method of vessel segmentation for X-ray coronary angiography images.* 2012. IEEE.

7. Felfelian, B., et al. *Vessel segmentation in low contrast X-ray angiogram images.* 2016. IEEE.

8. Vlachos, M. and E. Dermatas, *Multi-scale retinal vessel segmentation using line tracking.* Computerized Medical Imaging and Graphics, 2010. **34**(3): p. 213-227, 0895-6111.

9. Guerrero, J., et al., *Real-time vessel segmentation and tracking for ultrasound imaging applications.* IEEE transactions on medical imaging, 2007. **26**(8): p. 1079-1090, 0278-0062.

10. Dehkordi, M.T., et al., *Local feature fitting active contour for segmenting vessels in angiograms.* IET Computer Vision, 2013. **8**(3): p. 161-170, 1751-9640.

11. Nasr-Esfahani, E., et al. *Vessel extraction in X-ray angiograms using deep learning.* 2016. IEEE.

12. Nasr-Esfahani, E., et al., *Segmentation of vessels in angiograms using convolutional neural networks.* Biomedical Signal Processing and Control, 2018. **40**: p. 240-251, 1746-8094.

13. Otsu, N., *A threshold selection method from gray-level histograms.* IEEE transactions on systems, man, and cybernetics, 1979. **9**(1): p. 62-66, 0018-9472.

14. Ronneberger, O., P. Fischer, and T. Brox. *U-net: Convolutional networks for biomedical image segmentation.* 2015. Springer.

15. Melinščak, M., P. Prentašić, and S. Lončarić. *Retinal vessel segmentation using deep neural networks.* 2015.

16. Pereira, S., et al., *Brain tumor segmentation using convolutional neural networks in MRI images.* IEEE transactions on medical imaging, 2016. **35**(5): p. 1240-1251, 0278-0062.

17. Saeedan, F., et al. *Detail-preserving pooling in deep networks.* 2018.

18. Szegedy, C., et al. *Rethinking the inception architecture for computer vision.* 2016.

19. Tustison, N.J., et al. *N4ITK: improved N3 bias correction with robust B-spline approximation.*

20. Yan, Z., X. Yang, and K.-T. Cheng, *Joint segment-level and pixel-wise losses for deep learning based retinal vessel segmentation.* IEEE Transactions on Biomedical Engineering, 2018. **65**(9): p. 1912-1923, 0018-9294.

21. Kong, T.Y. and A. Rosenfeld, *Topological algorithms for digital image processing.* 1996: Elsevier.

22. Maurer, C.R., R. Qi, and V. Raghavan, *A linear time algorithm for computing exact Euclidean distance transforms of binary images in arbitrary dimensions.* IEEE Transactions on Pattern Analysis and Machine Intelligence, 2003. **25**(2): p. 265-270, 0162-8828.

23. Kovesi, P.D., *MATLAB and Octave functions for computer vision and image processing.* Centre for Exploration Targeting, School of Earth and Environment, The University of Western Australia, available from: http://www. csse. uwa. edu. au/~ pk/research/matlabfns, 2000. **147**: p. 230.

24. Arbab-Zadeh, A. and J. Hoe, *Quantification of coronary arterial stenoses by multidetector CT angiography in comparison with conventional angiography: methods, caveats, and implications.* JACC: Cardiovascular Imaging, 2011. **4**(2): p. 191-202, 1936-878X.

25. Kingma, D.P. and J. Ba, *Adam: A method for stochastic optimization.* arXiv preprint arXiv:1412.6980, 2014.

26. Li, C.H. and C.K. Lee, *Minimum cross entropy thresholding.* Pattern recognition, 1993. **26**(4): p. 617-625, 0031-3203.

27. Yen, J.-C., F.-J. Chang, and S. Chang, *A new criterion for automatic multilevel thresholding.* IEEE Transactions on Image Processing, 1995. **4**(3): p. 370-378, 1057-7149.

28. Zack, G.W., W.E. Rogers, and S.A. Latt, *Automatic measurement of sister chromatid exchange frequency.* Journal of Histochemistry & Cytochemistry, 1977. **25**(7): p. 741-753, 0022-1554.

29. Sauvola, J. and M. Pietikäinen, *Adaptive document image binarization.* Pattern recognition, 2000. **33**(2): p. 225-236, 0031-3203.

30. Sabour, S., N. Frosst, and G.E. Hinton. *Dynamic routing between capsules*. 2017.

31. Neumann, F.-J., et al., *2018 ESC/EACTS guidelines on myocardial revascularization.* European heart journal, 2019. **40**(2): p. 87-165.

32. Fihn, S.D., et al., *2012 ACCF/AHA/ACP/AATS/PCNA/SCAI/STS guideline for the diagnosis and management of patients with stable ischemic heart disease: a report of the American College of Cardiology Foundation/American Heart Association task force on practice guidelines, and the American College of Physicians, American Association for Thoracic Surgery, Preventive Cardiovascular Nurses Association, Society for Cardiovascular Angiography and Interventions, and Society of Thoracic Surgeons.* Journal of the American College of Cardiology, 2012. **60**(24): p. e44-e164, 0735-1097.

33. Maron, D.J., et al., *Initial invasive or conservative strategy for stable coronary disease.* New England Journal of Medicine, 2020. **382**(15): p. 1395-1407, 0028-4793.

34. Bober, R.M., et al., *The impact of revascularization on myocardial blood flow as assessed by positron emission tomography.* European journal of nuclear medicine and molecular imaging, 2019. **46**(6): p. 1226-1239, 1619-7070.

35. Gould, K.L., et al., *Mortality Prediction by Quantitative PET Perfusion Expressed as Coronary Flow Capacity With and Without Revascularization.* JACC: Cardiovascular Imaging, 1936-878X, 2020.